\documentclass[10pt,journal,compsoc,twoside]{IEEEtran}
\usepackage{comment}
\usepackage[normalem]{ulem}

\usepackage{tabularx,booktabs}
\newcolumntype{C}{>{\centering\arraybackslash}X} 
\usepackage{lipsum}

\ifCLASSOPTIONcompsoc
  \usepackage[nocompress]{cite}
\else
  \usepackage{cite}
\fi
\ifCLASSINFOpdf
   \usepackage[pdftex]{graphicx}
   \graphicspath{{authors/}{figures/}{./}}
   \DeclareGraphicsExtensions{.pdf,.png,.jpg,.jpeg}
\else
   \usepackage[dvips]{graphicx}
   \graphicspath{{eps/}}
   \DeclareGraphicsExtensions{.eps}
\fi
\usepackage{xcolor,colortbl,soul}

\colorlet{red}{black}
\colorlet{blue}{black}
\colorlet{green}{black}
\colorlet{purple}{black}
\colorlet{orange}{black}
\colorlet{cyan}{black}

\usepackage{url} 
\usepackage{soul}  
\usepackage{tcolorbox} 
\usepackage{paralist}
\usepackage{makecell}
\usepackage{multirow} 
\usepackage{siunitx}
\usepackage{amsmath,amssymb}   
\DeclareFontFamily{U}{cal}{}
\DeclareFontShape{U}{cal}{m}{n}{<->cmsy10}{}
\DeclareSymbolFont{rcal}{U}{cal}{m}{n}
\DeclareSymbolFontAlphabet{\mathcal}{rcal}

\usepackage{algorithm,algcompatible,amsmath}

\algnewcommand\INPUT{\item[\textbf{Input:}]}%
\algnewcommand\OUTPUT{\item[\textbf{Output:}]}%

\usepackage{cite}                      
\usepackage{tabu}                      
\usepackage{booktabs}                  

\usepackage{etoolbox}
\patchcmd{\quote}{\rightmargin}{\leftmargin 0.5em \rightmargin}{}{}

\usepackage{amsmath}
\newcommand{\addition}[0]{}

\definecolor{TODOcolor}{HTML}{fa9fb5}

\definecolor{FANcolor}{HTML}{dd3497}

\begin{document}
%
\title{GeoLinter: A Linting Framework for\\ Choropleth Maps}
%
%
%
%

\author{
        Fan Lei,
        Arlen Fan,
        Alan M. MacEachren, 
        Ross Maciejewski
\IEEEcompsocitemizethanks{
    \IEEEcompsocthanksitem 
F. Lei, A. Fan, and R. Maciejewski are with the School of Computing, Informatics \& Decision Systems Engineering, Arizona State University. E-mail: \{flei5,afan5,rmacieje\}@asu.edu.
\IEEEcompsocthanksitem A.M. MacEachren is with Pennsylvania State University. E-mail: maceachren@psu.edu.
    }
\thanks{Manuscript received MM DD, 2022; revised MM DD, 2022.}
}

%
%

\markboth{IEEE Transactions on Visualization and Computer Graphics,~Vol.~XX, No.~X, Month~YYYY}
{Lei \MakeLowercase{\textit{et al.}}: GeoLinter: A Linting Framework for Choropleth
Maps}
%



\IEEEtitleabstractindextext{%
\begin{abstract}

Visualization linting is a proven effective tool in assisting users to follow established visualization guidelines. Despite its success, visualization linting for choropleth maps, one of the most popular visualizations on the internet, has yet to be investigated. In this paper, we present GeoLinter, a linting framework for choropleth maps that assists in creating accurate and robust maps. Based on a set of design guidelines and metrics drawing upon a collection of best practices from the cartographic literature, GeoLinter detects potentially suboptimal design decisions and provides further recommendations on design improvement with explanations at each step of the design process. We perform a validation study to evaluate the proposed framework's functionality with respect to identifying and fixing errors and apply its results to improve the robustness of GeoLinter. Finally, we demonstrate the effectiveness of the GeoLinter - validated through empirical studies - by applying it to a series of case studies using real-world datasets.
\end{abstract}

}

\maketitle

\IEEEdisplaynontitleabstractindextext

%
\IEEEpeerreviewmaketitle

\IEEEraisesectionheading{\section{Introduction}\label{sec:introduction}}
\IEEEPARstart{T}{he} rise in visualization popularity has coincided with the rise in easy to use programming languages and software that have opened the design door to novices and experts alike. This has led to data visualizations becoming an extremely popular means of sharing and explaining data. However, the ease of creating visualizations has also enabled the creation of ill-designed visualizations as inexperienced designers who lack knowledge of visualization principles may create graphics that are poorly formed. To help overcome this knowledge gap, recent work in the visualization community has popularized the idea of visualization linting~\cite{chen2021vizlinter,hopkins2020visualint,mcnutt2018linting} where potential design issues are highlighted in the same manner as coding errors using a linting feature that constantly checks to see if the output follows design guidelines. When a violation of a design guideline occurs, the linting tools suggest corrections and alternatives to potentially improve the design.

Given the frequency of visualization construction errors~\cite{hopkins2020visualint}, researchers have begun to develop tools that identify and fix~\cite{chen2021vizlinter} malformed charts. Recent work has developed tools to recommend visualizations based on statistical properties of the data~\cite{wongsuphasawat2015voyager} or how the visualization is used~\cite{duke2001modular}. There is also an automated approach for producing various types of visualizations~\cite{wills2010autovis}, capable of generating graphics without manual definitions. Unfortunately, these tools~\cite{hopkins2020visualint, chen2021vizlinter} only cater to simple chart types (\textit{e.g.} line, bar, pie chart), missing one of the most common visualizations: the choropleth map. Research has shown that upwards of 30\% of D3 visualizations on the internet are choropleth maps, making them the most used visualization type in the D3 category~\cite{battle2018beagle}. Given the prevalence of choropleth maps, we believe that the visualization research and practice community can benefit from a framework for linting choropleth maps.

In this paper, we propose GeoLinter, a choropleth map linting framework. GeoLinter borrows concepts from previous linting tools~\cite{hopkins2020visualint, chen2021vizlinter} and expands on these concepts to support the design of choropleth maps. We first define a set of best practices and performance measures and then take an interactive approach in detecting and correcting suboptimal designs. Due to the numerous design considerations that go into creating a choropleth map (\textit{e.g.} the projection, classification method, color selection), GeoLinter divides the design process by first detecting encoding errors and then providing guidance on design choices. We note that GeoLinter is intended to help designers without cartographic experience avoid blunders, rather than to support novel mapping strategies.

Visualization linters are meant to assist in creating guideline-abiding visualizations that adhere to accepted design guidance. While other linters may exist~\cite{hopkins2020visualint, chen2021vizlinter}, they only cater to simple chart types such as bar and line charts. GeoLinter serves to expand on visualization linting by adding support for choropleth maps. By promoting a solid understanding of cartographic design principles among designers, it helps ensure the creation of visualizations that conform to essential rules. This is important because there are many visualizations that do not conform to guidelines in both the academic and professional community (\textit{e.g.} VisLies), and visualization software systems often have rulebases for information visualization broadly, but not for \addition{choropleth maps or any other thematic map types} specifically~\cite{chen2021vizlinter,hopkins2020visualint,mcnutt2018linting}. As such, our work is of significant relevance to the broad visualization community, with the target audience including visualization practitioners without cartography training.


To demonstrate our approach, we implement the proposed GeoLinter for VegaLite-based specifications and illustrate its effectiveness through a variety of case studies. We carried out a study to gather insights and observations on the maps produced by our target audience. Based on the results and feedback from an expert cartographer, we refined GeoLinter to enhance its reliability. Our contributions include:
\begin{itemize}
    \item GeoLinter, a framework for detecting, annotating, and revising potential design flaws in choropleth maps;
    \item Guidelines for design parameters in choropleth maps, informed by cartographic literature, and;
    \item A validation study exploring the functionality of GeoLinter in linting and education.
\end{itemize}

\section{Related Work}
\label{sec:related_works}

In this section, we summarize related works on visualization linting and recommendation tools.

\subsection{Visualization Linting}
Originally, ``linters'' were for the automated checking of source code for syntax and stylistic errors. Over the years, researchers saw a need to extend linting capabilities beyond just programming languages. Before its use in visualization, linting tools were developed for data in tabular form. CheckCell~\cite{barowy2014checkcell} and ExcelLint~\cite{barowy2018excelint} use an outlier and anomaly detection approach to find potential errors in spreadsheet entries. Later, with the onset of large databases, ConTest~\cite{mucslu2015preventing} was proposed for continuous-data-testing, preventing errors in large databases. Due to the importance of data preprocessing in creating robust visualizations, this work set the stage for later research in visualization linting.

McNutt and Kindlmann~\cite{mcnutt2018linting} were one of the first to introduce a visualization linter. Their work defines a set of rules that consider the aesthetics and readability of a visualization and outputs a binary decision with respect to the feasibility of \addition{various simple VegaLite~\cite{satyanarayan2016vega} chart types such as bar charts and line charts}. VisuaLint~\cite{hopkins2020visualint} flags visualization errors by drawing a red squiggly line over the location of the error, similar to how Microsoft Word sketches a red line underneath a misspelled word. VizLinter~\cite{chen2021vizlinter} accepts a VegaLite specification as input and translates it into answer set programming (ASP) facts to detect violated rules. It then proceeds to use a fixer engine to correct the rules to output a corrected specification. While our work draws ideas and design considerations from previous linting systems, the focus is on extending linting to choropleth maps, which, to date, remain unsupported in visualization linting. 


\subsection{Visualization Recommendation}

Visualization recommendation systems are another approach to supporting the visualization design process, where visualization recommendation systems attempt to compute (and recommend) a visualization that adheres to best principle design practices. Rule-based recommender systems pick a layout by following a set of rules defined by domain experts. VizDeck~\cite{perry2013vizdeck} is one such system that accomplishes this by suggesting a set of appropriate visualizations given the statistical properties of the data. Voyager~\cite{wongsuphasawat2015voyager} takes a similar approach as VizDeck but instead emphasizes browsing, which facilitates exploratory tasks.

Later, hybrid approaches were developed, which combine learning and manually defined rulesets for visualization. For example, Moritz et al.~\cite{moritz2018formalizing} propose Draco, a constraint-based system that learns to assign a score to visualizations by rewarding charts that follow graphical perceptual principles and penalizing rule violations. Dziban~\cite{lin2020dziban} expands on Draco by adding chart similarity logic, enabling chart recommendations to remain similar to an anchor chart, making the results more predictable. Finally, recent research has led to the development of machine learning based systems. Qian et al.~\cite{qian2020ml} pioneered the first machine learning based visualization recommender system trained on a large corpus of datasets and visualizations, \addition{which supports primitive chart types such as line charts, bar charts, and scatterplots.}
Li et al.~\cite{li2021KG4Vis} propose KG4Vis, a system that uses a knowledge graph to address the black-box nature of machine learning based recommender systems in an attempt to make the results more explainable.  However, these recommendation systems focus on simple chart types \addition{(\textit{e.g.} bar charts, box plots, heatmaps, histograms, line charts and scatter plots)}, with no support for choropleth maps.



\section{Choropleth Map Design}
\label{sec:background}
In this section, we outline all of the components necessary to create a choropleth map. We discuss design considerations, visualization grammar, and finally, our design goals.

\subsection{Design Considerations}
\label{subsection:design_considerations}
Choropleth maps are maps that display statistical data for \addition{enumeration units} (\textit{e.g.}, states, counties) typically using an \addition{sequential color scheme} to shade each area to reflect the class into which each area's data value fits~\cite{dent1999cartography}. In Slocum et al.'s~\cite{slocum2014thematic} textbook on Cartography and Geographic Visualization, the authors \addition{outline major topics for choropleth maps}: \addition{data classification, factors for selecting a color scheme, and details of color specification.}
We condense the map elements into three major design considerations: (1) classification \addition{method}, (2) symbology, and (3) projection.


\vspace{1.3mm} \noindent \textbf{(C1) Classification \addition{Method}}:
Before applying a data classification \addition{method}, the designer must choose the number of classes, or unique shades of color, in the map~\cite{cromley1995classed}. The choice of classification \addition{method} is a decades-long debate that ultimately affects the visual impression and reader interpretation of the map~\cite{evans1977selection}. An improper choice of the number of classes or data classification method may fail to portray key trends and patterns that would otherwise be visible. For simple maps, cartographers generally recommend that the number of classes is confined between three (3) to seven (7)~\cite{slocum2014thematic}. \addition{This recommendation is in line with Miller's observation on the magic number 7±2, suggesting that there are cognitive limits on our capacity for processing distinct chunks of information, which is a foundational guideline in cartography\cite{miller1956magical}.} Evans~\cite{evans1977selection} recommends that five or fewer classes be used for novice map readers. ColorBrewer~\cite{harrower2003colorbrewer} supports 11 colors for diverging schemes and 9 colors for sequential schemes, all of which are colorblind safe. Thus, our framework suggests classed maps between 3-11 classes. Anything above 11 classes we consider to be an unclassed map, which maps the range of the data to a continuous spectrum of colors. 
Classification \addition{methods} divide rank-ordered data into a specific number of bins. In choropleth maps, these bins are then mapped to different shades of a selected color scheme. Zhang and Maciejewski~\cite{zhang2016quantifying} identify the problem that data values can be close to a classification boundary and a slight change in the classification \addition{method} can result in different clusters. Cartographers have studied the performance of different classification strategies under varying use cases and data distributions. Equal intervals, which partitions the overall data range into intervals of equals-sized ranges, are best suited for uniform distributions~\cite{evans1977selection} but can have inconsistent or unpredictable performance~\cite{smith1986comparing}. Mean standard deviation is best used when the distribution is unimodal~\cite{evans1977selection}. Quantiles are one of the most versatile classification methods, suited for normal distributions~\cite{smith1986comparing}. Our framework uses properties of the data to recommend an appropriate classification.

\vspace{1.3mm} \noindent \textbf{(C2) Color and Symbology: }In a choropleth map, the symbolic features refer to the fill colors, \addition{outline width}, and text. Each of these features can directly influence the choropleth map's readability~\cite{brychtova2016empirical}. For example, choosing two fill colors that are too similar to each other may make comparison tasks difficult~\cite{brychtova2015discriminating,brychtova2017effect}. Also, the choice of \addition{outline width} for area boundaries can influence the degree to which overall patterns versus individual values are emphasized. \addition{A narrow stroke or no stroke for boundaries enhances overall patterns, but completely omitting the stroke might not be a good solution since it could result in the loss of essential information; when neighboring units have the same color hue, the absence of a delineating stroke can cause them to visually merge, hiding the information of individual enumeration units. Bold strokes direct attention to individual enumeration areas.} The border frame should also be appropriately selected.

Color choice is crucial in creating choropleth maps. ColorBrewer~\cite{harrower2003colorbrewer} is a tool designed to support appropriate color palette selections for choropleth maps. Brychtová and Çöltekin~\cite{brychtova2015discriminating} find that the more distinct colors are in choropleth maps, the faster readers are able to distinguish and complete comparison tasks. This effect was confirmed by a controlled, in-lab study with an eye-tracker~\cite{brychtova2016empirical}. Apart from \addition{color discriminability}, other perceptual features of choropleth colors must also be taken into account. For example, red regions are perceived as having more area than green ones~\cite{cleveland1983color}. \addition{Notably, a foundational principle in designing sequential color schemes is that people tend to associate higher values with darker colors~\cite{schiewe2019empirical,mcgranaghan1989ordering,association2006geographic}.} It follows that if the choropleth needs to show logical ordering of values, then rainbow colors must be avoided~\cite{golebiowska2020rainbow}. Thus, in GeoLinter, the palettes recommended are standard sequential and diverging palettes recommended by ColorBrewer~\cite{harrower2003colorbrewer}.

\vspace{1.3mm} \noindent \textbf{(C3) Projection:} Creating a map requires transforming from the sphere to a 2D plane leading to a variety of distortions. A popular method of quantifying distortion in map projection is the use of Tissot's indicatrix~\cite{laskowski1989traditional}. By projecting a circle from the spherical earth onto a two-dimensional map, the reader can see what kind of distortion occurs. Common map projections often take one of two extremes: (a) equal area – which preserves correct area everywhere but must distort angular relations around points, and (b) conformal – which preserves angular relations around points but must distort area. Equal area projections are almost always preferable for \addition{choropleth maps} since it is important to represent the size of places correctly as readers try to understand spatial patterns~\cite{usery2001all}. Conformal projections were originally developed for situations in which angular and directional relations are important. The most common projection on the web is the Mercator projection, a conformal projection designed to map the entire globe while also drastically distorting both area and shape, thus making it unsuited for data mapping~\cite{Leon2020}. However, recent internet tools have used other map projections and have even facilitated education about map projections for ordinary audiences~\cite{kessler2017map}.
There are situations in which compromise projections are chosen (e.g., when some distortion of areas is acceptable in exchange for pleasing appearance at the global scale or rhetorical effect). At the global scale, however, the best choice is now considered by many to be an equal area projection inspired by the attractive Robinson global compromise projection – the Equal Earth projection, jointly developed by Šavrič et al.~\cite{vsavrivc2019equal}. In many cases, Šavrič et al.'s natural earth projections are also appropriate~\cite{vsavrivc2011polynomial} 
For regions smaller than the globe, the best projections for data maps are equal area projections designed specifically to minimize other distortion across the area of interest. For the USA, as an example, the Albers Conic Equal Area projection is appropriate as it minimizes size distortions of the map~\cite{snyder1978equidistant}.
For small scale maps much smaller than the scale of the Earth, conformal projections such as Mercator can be used, as long as the distortion of the area is minimal. 
The specific equal area projection that works best will depend on the extent of the geographic area depicted (from global, through continental and national, to more local) as well as on the specific location on the globe and the shape of the region (\textit{e.g.}, an Albers equal area projection with standard parallels positioned appropriately is good for east-west extending places in mid-latitudes like the USA while a Lambert azimuthal equal-area projection with the center point in the center of the country is good for an elongated north-south region like Argentina). Our framework considers the geographic extent and recommends an appropriate projection for the map with the support of a projection wizard proposed by Šavrič et al.~\cite{vsavrivc2016projection}.

\begin{table*}[tbh]
\renewcommand{\arraystretch}{1.4}
\begin{tabularx}{\textwidth}{@{}l*{10}{C}}
  \hline 
    \textbf{Name}  &
    \textbf{Description} &
    \textbf{Action}
         \\[0.5ex] 
  \hline
    \color[HTML]{C00000}{\textbf{DATA\_URL}}  &
    \textlangle{}data.url\textrangle{} must be nonempty. &
    The data file must be specified.
         \\[0.5ex] 
  \hline
    \color[HTML]{C00000}{\textbf{DATA\_FEATURE}}  &
    \textlangle{}data.format.feature\textrangle{} must be nonempty. &
    At least one geographical feature must be specified.
         \\[0.5ex]
   \hline
    \color[HTML]{C00000}{\textbf{MARK}}  &
    \textlangle{}mark\textrangle{} must have the value ``geoshape''. &
    For a choropleth map, the only viable option for \textlangle{}mark\textrangle{} is ``geoshape'' in VegaLite.
         \\[0.5ex]
  \hline
    \color[HTML]{C00000}{\textbf{COLOR\_FIELD}}  &
    \textlangle{}encoding.color.field\textrangle{} must be nonempty. &
    For a choropleth map, the color field must not be empty.
         \\[0.5ex]
  \hline
    \color[HTML]{C00000}{\textbf{COLOR\_TYPE}}  &
    \textlangle{}color.type\textrangle{} should have the value ``quantitative'' for a choropleth map or ``nominal'' for a thematic map that portrays regions of categorical data. Nominal values should use categorical colors. &
    The \textlangle{}color.type\textrangle{} must be either quantitative or nominal.
         \\[0.5ex]
  \hline
    \hline
\end{tabularx}

\begin{tabularx}{\textwidth}{@{}l*{10}{C}c@{}}
  \hline
    \color[HTML]{FFC000}{\textbf{NUM\_CLASSES}}  &
    The number of classes is outside of the recommended range 3-7. Unclassed maps are not allowed. &
    Display the top performing classification methods for 3-7 classes, sorted by GVF or Moran's I.
         \\[0.5ex]
  \hline
    \color[HTML]{FFC000}{\textbf{LEGEND\_COLOR}}  &
    The legend contains 2 or more colors that are too similar. &
    Suggest an array appropriate color palette for the legend.
         \\[0.5ex]
  \hline
    \color[HTML]{FFC000}{\textbf{BORDER\_COLOR}} &
    The border color is too similar to the legend colors. &
    Suggest an appropriate color palette for the legend, and reset the border color to black.
         \\[0.5ex]
  \hline
    \color[HTML]{FFC000}{\textbf{BG\_COLOR}} &
    The background color is too similar to the border or legend colors &
    Suggest an appropriate color palette for the legend, and set the background color to white.
         \\[0.5ex]
  \hline
    \color[HTML]{FFC000}{\textbf{LOW\_GVF}} &
    The classification GVF score is low. This is triggered if the current GVF is lower than the average GVF of all possible classification methods across all number of classes. &
    Display the top performing classification methods for 3-7 classes, sorted by GVF or Moran's I.
         \\[0.5ex]
  \hline
    \color[HTML]{FFC000}{\textbf{PROJ*}} &
    The map projection is not appropriate for the given geographic extent. &
    Based on the geographic extent from GeoJSON, recommend the most appropriate projection with the support of a projection wizard~\cite{vsavrivc2016projection}.
         \\[0.5ex]
  \hline

    \color[HTML]{FFC000}{\textbf{DATA\_NORM*}} &
    The map should portray a normalized value and not an absolute value. &
    Suggest data normalization techniques as recommended by Slocum et al.~\cite{slocum2014thematic}
         \\[0.5ex]
  \hline

      \color[HTML]{FFC000}{\textbf{TITLE\_LEGEND*}} &
    The legend should include the data units being mapped. The title should be present and preferably reference the map theme, location, and time being mapped. &
    Automatically label the legend using the units from the data normalization stage or a custom-defined label. Generates a title based on the units and enforces a non-empty title.
         \\[0.5ex]
  \hline
\end{tabularx}

 \vspace{2mm}
 \caption{The list of rule violations detected and corrected by GeoLinter. Encoding errors are listed above the double horizontal line and soft rules below. Rules marked by an asterisk were added in or refined after conducting a validation study (Section~\ref{sec:evaluation}).}
 \vspace{-2mm}
 \label{tab:rules}
\end{table*}

\begin{figure*}[tbh]
	\centering
	\includegraphics[width=2.0\columnwidth]{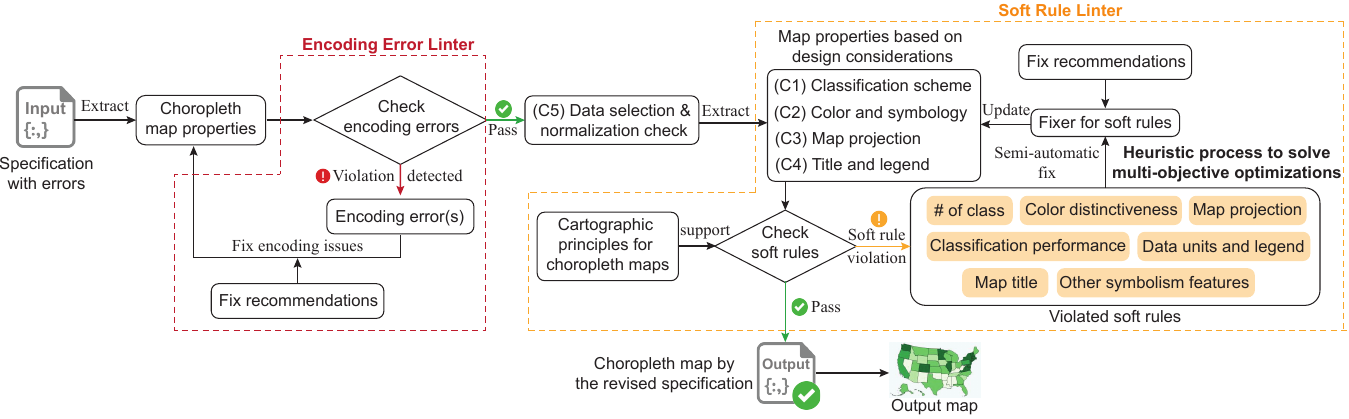}
	\vspace{-2.9mm}
  \caption{The end-to-end pipeline of GeoLinter. The soft rule linter is the foundation of our framework, which offers guidance on design parameters based on best practices in cartographic literature.}
  \vspace{-4mm}
 \label{fig:pipeline}
\end{figure*}

\subsection{Grammar Considerations}
\label{subsection:grammar_considerations}

Along with considerations directly related to choropleth map design, grammar and syntax must also be considered. While our framework is agnostic to grammar or encoding languages, we have chosen to instantiate our work using VegaLite~\cite{satyanarayan2016vega}, a high-level visualization grammar known for its conciseness and use in other extant linting tools~\cite{hopkins2020visualint, chen2021vizlinter}. Within VegaLite, marks and encoding channels must be explicitly declared. For choropleth maps, the options for mark and encoding channel are confined to geographical shape and color, respectively. Position, orientation, and size are also considered, but are less customizable. For example, maps are generally positioned at the center. North is usually pointed up, and size is adjusted so that the map can fit within the screen. Thus, declarative errors for choropleth maps are limited. However, due to the open-ended nature of the other design considerations (class-interval selection, symbology, and projection), solving these issues is a nontrivial problem. We introduce an engine that checks for required encoding prerequisites that are necessary for the choropleth map. Encoding errors refer to invalid grammar specifications that may lead to visualizations that cannot compile. We identify two major categories of encoding problems: 1) missing the key properties related to the map visualization in the specification, and 2) invalid property types and values for the choropleth map visualization.

\subsection{Design Goals}
\label{subsection:design_goals}

Given these design considerations, it is clear that the design of a choropleth map is a multi-objective problem, and the considerations may vary according to task. There is no absolutely ``correct'' choropleth map~\cite{Xiao2006}; however, there are multiple best practice design guidelines from the cartography community. We have identified three key design goals for building a linting framework to support choropleth map development. A summary of the guidelines can be found in Table~\ref{tab:rules}.

\vspace{1.5mm} \noindent \textbf{G1: Detect and indicate encoding errors.}
Encoding errors prevent the choropleth map from compiling or rendering. The first goal is to detect these errors in the grammar specification and indicate the location of the error along with an explanation as to why the map failed to render.

\vspace{1.5mm} \noindent \textbf{G2: Detect and explain suboptimal design parameters (C1, C2, C3).}
Poor choropleth map design can fail to portray important trends that need to be visualized. Our second goal is to identify any design choices (\textbf{C1: Classification, C2: Color and symbology, C3: Map projection}) that may be problematic. Not all visualization designers have experience in cartography, so we also support explaining domain-specific technical terms in an understandable way. We refer to the design parameters related to \textbf{C1}, \textbf{C2}, and \textbf{C3} as soft rules, and they are summarized in Table~\ref{tab:rules}.

\vspace{1.5mm} \noindent \textbf{G3: Automatically fix the detected errors (C1, C2, C3).}
Surfacing the error alone is inadequate. For example, just indicating that the classification method is problematic does not help a novice mapmaker the visualization designer, as they may not even be aware of different classification methods. Our framework provides options to correct the errors by calculating a set of good classification \addition{methods} and enabling their comparison. Mathematical definitions of our methods and metrics and a list of best practices are explained in-depth in Sections~\ref{subsubsec:classification}, ~\ref{subsubsec:color_and_symbology}, and~\ref{subsubsec:map_projection}.

\vspace{1.5mm} \noindent \textbf{G4: Support integration to GIS frameworks.} A geographical information system (GIS) is a database containing geographic data with software tools for analyzing and visualizing those data~\cite{chang2008introduction}. GIS has become a standard for geoinformatics in both industry and academia. Our framework is designed to support GeoJSON files, an accepted format for defining geographical shapes, along with their non-spatial properties. GeoJSON naturally lends itself for automatic integration into existing GIS frameworks, strengthening the reach of our framework.

\section{The GeoLinter Framework}
\label{sec:framework}
\begin{figure*}[tbh]
	\centering
    \includegraphics[width=\linewidth]{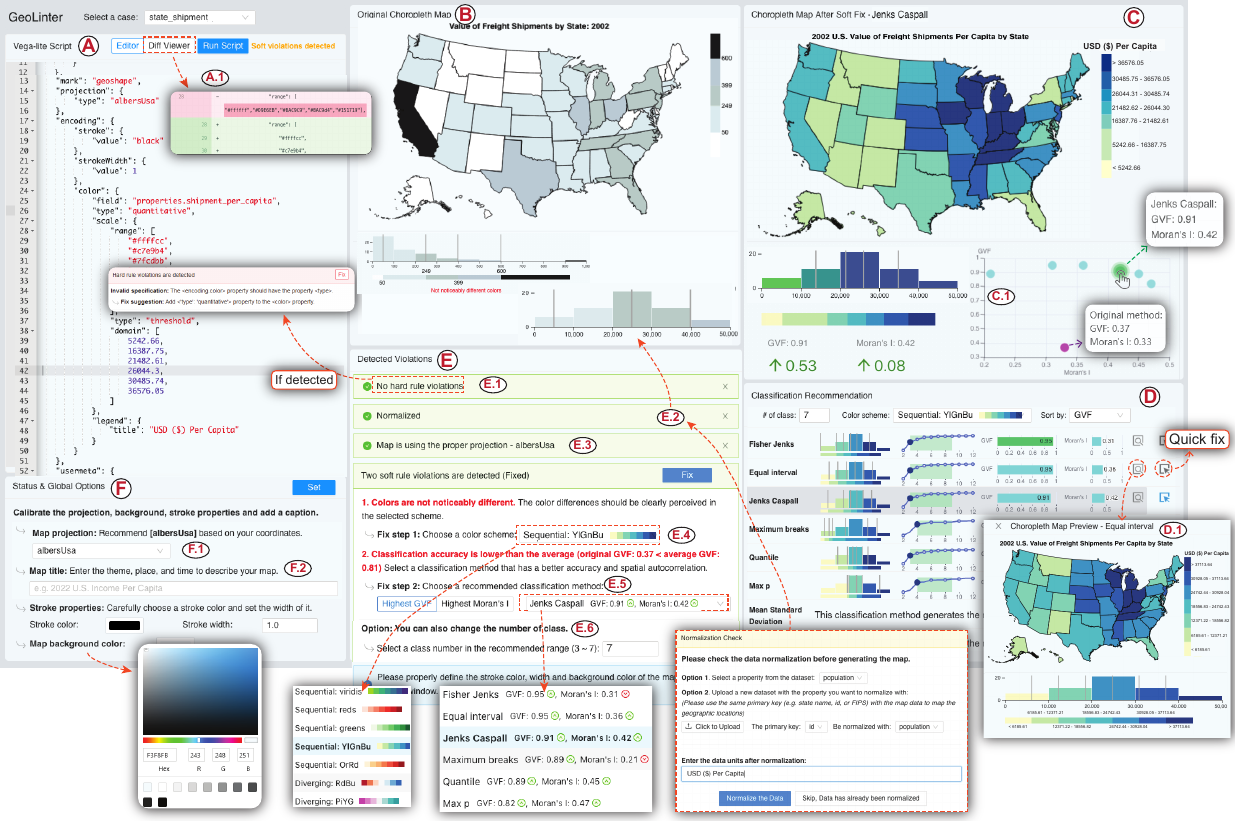}
    \caption{The GeoLinter Interface: (A) the VegaLite code editor; (B) the original map view; (C) the map after applying soft fixes; (D) classification recommendations based on GVF and Moran's I; (E) detected violations with guides and prompts on map design improvements, and; (F) the status and global options panel. A choropleth map showing the value per capita of freight shipments in the U.S. by state 2002. In the original choropleth map design (B), the data classification accuracy is lower than the average value; the colors between bins are nearly indistinguishable; the map data has not been normalized and the data units are missing. After applying the suggested fixes from GeoLinter, the designer produces (C).}
    \vspace{-3mm}
    \label{fig:teaser}
\end{figure*}

Based on the design goals, we have developed GeoLinter (Figure~\ref{fig:pipeline}), a linting framework for choropleth maps. GeoLinter is based on the theory of choropleth map design and lints based on derived guidelines. A series of encoding errors and soft rules form the basis of our linting procedure. Table~\ref{tab:rules} summarizes all of the rules in GeoLinter and the workflow is documented in Figure~\ref{fig:pipeline}. Figure~\ref{fig:teaser} shows the interface of the framework.

To create our ruleset, we designed specifications to create a viable choropleth map using visualization grammar. We developed a set of encoding rules to satisfy goal \textbf{G4}. Five properties serve as a prerequisite to making a choropleth map within VegaLite: (1) the dataset must be specified; (2) at least one geographical feature must be specified; (3) the \textlangle{}mark\textrangle{} property must be ``geoshape''; (4) the \textlangle{}encoding.color.field\textrangle{} property must be nonempty, and; (5) the \textlangle{}color.type\textrangle{} property must be either ``quantitative'' or ``nominal''. The fifth rule enables designers to visualize both statistical and categorical data. We also highlight that our system supports ordinal data, which possesses an inherent order without a quantifiable distance between data points~\cite{kraak2020cartography}. Since encoding errors prevent a choropleth map from rendering properly, resolving this step is mandatory. While other design parameters (\textit{e.g.} color, background) are also required, VegaLite has predefined default values if they are not explicitly specified. Our framework automatically collects the value based on the VegaLite defaults. Here, our hard rules are functions of the grammar syntax and not design. We classify our design guidelines as soft rules and define them over a set of three guidelines to satisfy our design constraints and goals: classification guidelines \textbf{C1, G2, G3}, symbology guidelines \textbf{C2, G2, G3}, and map projection guidelines \textbf{C3, G2, G3}. 


We first conceptualized and developed the initial framework of GeoLinter, aiming to help visualization designers prevent common mistakes in choropleth map creation. This initial iteration of the tool was created based on established cartographic guidelines and principles, as detailed in this section. Following the establishment of this initial version of GeoLinter, we initiated an evaluation study (Section~\ref{sec:evaluation}). The purpose of this study was to gather feedback and to assess how effectively GeoLinter was assisting in reducing common design errors. The evaluation study's methodology and its key findings are detailed. During the evaluation study, we identified certain limitations in our tool as well as areas where designers were still creating erroneous maps, indicating areas for potential improvement. To gain deeper insights, we not only relied on the feedback from the evaluation study but also incorporated the valuable inputs from our co-author, a cartography expert, and the reviews from the initial draft of this paper. \addition{Based on the feedback, we refined GeoLinter, enhancing its interface and updating guidelines, \textit{detailed in Section~\ref{sec:geolinter_v2}}.}

\subsection{Classification Guidelines}
\label{subsubsec:classification}

Numerous classification methods for choropleth map design exist. We instantiate the most popular methods and define our linting recommendations on these methods based on the cartographic literature.

\vspace{1.3mm} \noindent \underline{Equal Intervals} - Divides attribute values into ranges of equal size~\cite{evans1977selection}. For a given number of classes $k$, equal intervals classification divides the classes into intervals having a width $w = \frac{x_0 - x_{n-1}}{k}$, with the first class being $(-\infty,x_{n-1}-(k-1)w]$, and the last class being $(x_{n-1}-w,x_{n-1}]$, where $x_0, x_{n-1}$ are the first and last entries in the dataset in sorted order. Although simple to construct and interpret, the result may include sparse or empty classes under datasets with extreme outliers~\cite{brewer2002evaluation}.

\vspace{1.3mm} \noindent \underline{Quantiles} - Distributes a set of values into classes that have approximately $|\frac{n}{k}|$ members~\cite{evans1977selection}. No classes are left empty or contain an unbalanced number of values. While quantiles avoids the problem of sparse classes, the intervals may be significantly different in range size, which can cause difficulties in interpretation~\cite{smith1986comparing}. Additionally, readers tend to assume that all places assigned to any given class are similar, but for skewed distributions, it is typical for at least one of the classes to contain values that are extremely different from one another.

\vspace{1.3mm} \noindent \underline{Mean Standard Deviation} - Uses the sample mean ($\Bar{x}$) and standard deviation ($s$) to define class boundaries as distance from the mean~\cite{evans1977selection}. The distance is usually a multiple of the standard deviation. A common setup for this method is to assign $k=5$, where the classes $c$ are defined as $c_i = \Bar{x} + (2-i)s, i \in \{0,1,2,3,4\}$. This classifier is best used when the data is normally distributed~\cite{evans1977selection}.

\vspace{1.3mm} \noindent \underline{Maximum breaks} - Considers the difference between sorted values to place breakpoints~\cite{stern2011statistics}. It considers how far apart each value is from the next one in the sorted sequence, then places $k-1$ breaks between the values that are furthest apart from each other. Maximum breaks is an appropriate method to use when the objective is to partition the data points in each class such that they are separated from those in adjacent classes. It performs well when the distribution is not unimodal. This method is deterministic given $k$, as running this algorithm will always produce the same result under the same dataset. However, because this method only considers the top $k-1$ differences between sorted values, within-group properties are often ignored~\cite{stern2011statistics}.

\vspace{1.3mm} \noindent \underline{Head tail} - This algorithm~\cite{jiang2013head} recursively partitions the data around iterative means until the distributions in each of the classes are no longer heavy-tailed. This method proves effective with heavy-tailed distributions, such as log-normal and power law distributions~\cite{jiang2013head}.

\vspace{1.3mm} \noindent \underline{Jenks\addition{-}Caspall} - \addition{Often considered a refined version of the original Jenks Natural Breaks method, the Jenks-Caspall classification seeks to minimize deviations around class means using a heuristic process~\cite{jenks1971error}. While the Jenks Natural Breaks method focuses on finding "natural" clusters in data sets, aiming to reduce variance within classes and maximize variance between them, the Jenks-Caspall method introduces a more systematic approach to this objective.}

\vspace{1.3mm} \noindent \underline{Fisher\addition{-}Jenks} - A dynamic programming approach to minimize the sum of deviations around class medians, loosely based on Jenks\addition{-}Caspall~\cite{jenks1971error}. Whereas Jenks\addition{-}Caspall may not provide an optimal classification for $k$ classes, Fisher\addition{-}Jenks is guaranteed to produce an optimal result.

\vspace{1.3mm} \noindent \underline{Max-p} - A classifier that uses the max-p~\cite{duque2012maxp} algorithm applied to map classification. In short, the max-p-regions algorithm aggregates $n$ areas into a number of homogeneous regions, such that they satisfy a contiguity constraint while maximizing intraregional homogeneity. Since it is a heuristic, there is no guaranteed optimal solution.


\vspace{1.3mm} \noindent \textbf{Classification Performance Measures} To determine the best choice of classifier, we utilize Goodness of Variance Fit (GVF) and Moran's I, two numerical measures of classification results in choropleth map design correlated with perceptual accuracy~\cite{brewer1997mapping,declerq1995choropleth}.

This is important, as relying solely on appearances and preferences when designing maps does not correlate with the accuracy and robustness of the map. Hegarty et al.~\cite{hegarty2009naive} found that individual preferences, even those of domain experts, are not a good indicator of display effectiveness. They stress the importance of conducting additional objective measures of display efficiency. In a study by Brewer and Pickle~\cite{brewer2002evaluation}, quantile classification resulted in the highest accuracy for basic analytical tasks such as data extraction and comparison. Jenks's natural breaks, equal interval performed satisfactorily, while methods such as shared area and box plot were significantly worse. Thus, the poorer performing class schemes were not used in our work, as separate studies have confirmed that they are perceptually less effective.

\vspace{1.3mm} \noindent \underline{GVF} - Smith's~\cite{smith1986comparing} GVF is defined as

\begin{equation}
    GVF = 100 - ((SSW/SST) \times 100),
\end{equation}

\noindent where \textit{SST} (Sum of Squares Total) is the sum of deviations squared of individual data values from the overall mean, and \textit{SSW} (Sum of Squares Within) is the sum of squared deviations of class data values from each class mean and then summing it for all classes. GVF is the preferred metric, and GeoLinter will mark the soft rule violation \textcolor[HTML]{FFC000}{\textbf{LOW\_GVF}} if the GVF is lower than the average GVF of all possible classification methods across all number of classes. Since all values of GVF were calculated, the threshold for GVF can be set at a certain percentile threshold. For the purposes of GeoLinter being a proof-of-concept, we arbitrarily set the threshold as the average GVF.

\vspace{1.3mm} \noindent \underline{Moran's I} - When considering spatial correlation, Moran's I~\cite{moran1950notes} is one of the most prominent measures, defined as:

\begin{equation}
    I = \frac{N}{\sum_i\sum_j w_{ij}}\frac{\sum_i \sum_j w_{ij}(x_i-\Bar{X})(x_j-\Bar{X})}{\sum_i (x_i-\Bar{X})^2},
\end{equation}

\noindent where $N$ is the number of spatial regions, $i,j$ are the indices of the regions, $x$ is the variable of interest, $\Bar{X}$ is the mean of $x$, $w$ is the matrix of spatial weights. The matrix $w$ can vary depending on how the weights are chosen and what definition of contiguity is used. In our framework, we use the Queen~\cite{rey2021pysal} definition of contiguity-based neighbors, which means that two regions are neighbors if and only if they share at least one point. The other definition of contiguity is Rook~\cite{rey2021pysal}, which counts two areas as neighbors if they share an edge. For irregular polygons (like most areal units encountered in practice), the differences between these definitions are slight. In order to deal with potential inaccuracies in the polygon file (such as rounding errors), using the queen criterion is recommended in practice~\cite{anselin2020contiguity}. Hence it is also the default for contiguity weights for polygon data in GeoDa~\cite{anselin2009geoda} and ArcGIS Pro~\cite{arcgis31docu}. We note that while useful to visualize the value of Moran's I, it is not a measure of classification performance, and it does not take into account maximizing interclass variances nor minimizing intraclass variances. The purpose of including this measure into our framework is to introduce an alternative metric that considers the spatial implications of the classification method without the more extreme solution of building spatial considerations into the classification formally.

\vspace{1.3mm} \noindent \textbf{Number of Classes} In our framework, the number of classes is confined to 3-7, conforming to classical cartographic recommendations~\cite{declerq1995choropleth}. Classes fewer than 3 are binary or single color maps, which oversimplify the map, while more than 7 classes may make color comparison and data extraction tasks difficult due to the overwhelming number of unique colors. A map that does not use the recommended range will trigger the soft rule violation \textcolor[HTML]{FFC000}{\textbf{NUM\_CLASSES}}. The GVF value also generally increases with the number of classes.  We calculate the recommended threshold value for the number of classes by starting at 2 classes and incrementing by 1 until the GVF surpasses 0.5. This value is based on finding a knee-point~\cite{Satopaa2011FindingA}, the point at which the cost to increase a parameter is no longer worth the increase in performance.

\subsection{Symbology Guidelines}
\label{subsubsec:color_and_symbology}

As a continuation of our soft rule linting, we detect and provide guidance on color and symbology, backed by cartographic literature.

\vspace{1.3mm} \noindent \textbf{Color Theory} When choosing a color palette, the main design consideration is to choose between sequential and diverging color palettes. We recommend diverging palettes when there is a meaningful point to contrast with the other values in the data~\cite{harrower2003colorbrewer}. For example, a map that spans both negatives and positives would likely use the diverging scheme, as zero would be the reference point. All other schemes should use a sequential color palette. Diverging and sequential schemes should be chosen on a per-task basis~\cite{tominski2008task}. For example, if a designer wants to emphasize below and above-average areas in the map, a diverging color scheme would be appropriate, where the average value would be the reference point. Ordinal data can also use sequential or diverging schemes, but categorical data should only use \addition{qualitative color scheme}. These recommendations are programmed as prompts into GeoLinter. We use various palettes from ColorBrewer~\cite{harrower2003colorbrewer}. ColorBrewer color palettes are optimized for white backgrounds. Due to the rise in popularity in dark backgrounds in interfaces, CARTO~\footnote{carto.com/carto-colors} supports colors against non-white backgrounds, using the foundational color schemes from ColorBrewer.

\vspace{1.3mm} \noindent \textbf{\addition{Color Discriminability}} \addition{Color discriminability} is primarily used for comparison tasks. Colors that are too similar to each other are not distinguishable by the human eye. ColorBrewer already includes schemes that have been empirically tested to be discriminable on maps.
We derive three rules based on \addition{color discriminability}: (1) \textcolor[HTML]{FFC000}{\textbf{LEGEND\_COLOR}}, which is violated when two colors from each class are too similar to be distinguishable, (2) \textcolor[HTML]{FFC000}{\textbf{BORDER\_COLOR}}, which is violated when the border color is too similar to any of the colors in the legend or the background, and (3) \textcolor[HTML]{FFC000}{\textbf{BG\_COLOR}}, which is violated when the background color is too similar to any of the colors in the legend or the border. Any of these errors are triggered when custom schemes barely meet the discriminability criterion. To determine if colors are too similar, we use a model by Stone et al.~\cite{stone2014engineering} that determines if two colored glyphs can be differentiated faithfully. We check each pairwise color combination in the legend to determine if this criterion is met. 

\vspace{1.3mm} \noindent \textbf{Symbology} Border \addition{outline width} is another design parameter that affects the interpretability of a map. Overly thick strokes emphasize the individual enumeration areas, while a narrow stroke enhances overall patterns. There are limited recommendations on \addition{outline width} from the cartographic literature, and GeoLinter simply enables manual adjustment of the border width. Finally, the interface also allows the adjustment of frame width and margins. 



\subsection{Map Projection Guidelines}
\label{subsubsec:map_projection}

The map projections supported by the GeoLinter framework include Albers (a conic equal area projection that minimizes shape distortion for features in the mid-latitudes and extending more east-west than north-south)~\cite{snyder1982map}; Albers USA~\cite{bostock2011d3} (a composite USA-centric projection configured for the lower 48 states, Hawaii, and Alaska); Equal Earth~\cite{vsavrivc2019equal}; Natural Earth (a compromise pseudocylindrical map projection that is designed to depict the entire earth while balancing area and shape distortion)~\cite{vsavrivc2011polynomial}, and; Mercator. Although the most popular projection on the web is the Mercator, it is not appropriate for choropleth maps due to its excessive area distortion. Based on the GeoJSON data, GeoLinter will list appropriate projections and will mark the soft rule violation \textcolor[HTML]{FFC000}{\textbf{PROJ}} if the current map projection is not an equal area projection. The best choice of projection for maps of the entire world is the Equal Earth projection due to its equal-area property combined with good handling of other distortion and pleasing appearance for global maps. Thus, it will be our default option. The Albers USA projections is recommended for USA maps, and attempts to minimize other kinds of distortion across the latitude-longitude range containing the USA. It is implemented as an illustration of how GeoLinter can handle maps for specific non-global places.

This guideline underwent a major change (see Table~\ref{tab:rules}) and is further discussed in Section~\ref{sec:geolinter_v2}.



\subsection{Implementation}

\addition{The frontend was implemented using ReactJS and D3~\cite{bostock2011d3} and consists of six panels, as illustrated in Figure~\ref{fig:teaser}. Each panel serves a unique purpose in helping to visualize the different elements and functionalities of GeoLinter.}

\label{subsubsec:interface}

\vspace{1.5mm} \noindent \textbf{VegaLite Script}: Figure~\ref{fig:teaser}A contains the VegaLite specification of the choropleth map. There is an editor and diff viewer panel (A.1) that tracks the changes that are made to the script. The reason for including the code view is for easy embedding into the web. For example, a designer can easily use the GUI interface to customize the map, while the code view automatically updates. The code for the final map is readily available for publishing.

\vspace{1.5mm} \noindent \textbf{Original Choropleth Map}: Figure~\ref{fig:teaser}B is a static image of the choropleth map that may or may not have soft rule violations. A histogram view of the original data distribution, class breaks, and colors used is provided.

\vspace{1.5mm} \noindent \textbf{After Soft Fix}: Figure~\ref{fig:teaser}C contains the final output of the map and mirrors the specification as defined by the VegaLite Script panel. Whereas the \textbf{Original Choropleth Map} view does not change, this view and the script are both updated whenever a fix is applied. A histogram is also shown here to assist in understanding the new class breaks. At the bottom left corner is a table detailing the change of GVF and Moran's I values. At the bottom right is a chart showing the GVF or Moran's I value of all possible classification methods. The original method is shown as a purple dot, and the current method is shown in green, allowing for a quick comparison of performance.

\vspace{1.5mm} \noindent \textbf{Classification Recommendation}: Figure~\ref{fig:teaser}D displays all of the classification methods (Section~\ref{subsubsec:classification}), given the number of classes, with their corresponding performance metric values. Prior to loading this, the GVF and Moran's I of all methods for all number of classes 3-11 were calculated using the PySAL~\cite{rey2021pysal} Python package, a library for geospatial data science. Each method displays a histogram, plots a line chart with the number of classes on the x-axis and GVF on the y-axis, and shows the raw values for GVF and Moran’s I. For each classification method used, a green box in the line chart indicates a range representing the recommended number of classes. Separately, we have organized the classification methods themselves in descending order, not based on the number of classes, but according to their respective GVF or Moran's I value.
There is also the option to preview the map and apply a quick fix (D.1).

\vspace{1.5mm} \noindent \textbf{Detected Violations}:
All detected encoding and soft rule violations are displayed in Figure~\ref{fig:teaser}E. We employ step-by-step narrative visual prompts to recommend actions. Any changes in this view will automatically update the VegaLite script and the After Soft Fix View. \addition{We note that GeoLinter carries out its rule evaluations independently. That is, each detected violation is evaluated separately. As an illustration, consider the case of low classification performance (\textcolor[HTML]{FFC000}{\textbf{LOW\_GVF}}). This violation is identified and evaluated independently of any others. While multiple approaches may address this issue - such as implementing a more appropriate classification method or adjusting the number of classes (\textcolor[HTML]{FFC000}{\textbf{NUM\_CLASSES}}) - each potential solution is regarded as a separate action and does not influence the evaluation of other violations or solutions.}

\vspace{1.5mm} \noindent \textbf{Status \& Global Options}: Figure~\ref{fig:teaser}F contains settings for the map projection, caption or title, border stroke settings, and background color. Any changes in this view will also update the VegaLite script and the After Soft Fix View.

\section{Evaluation}
\label{sec:evaluation}

\begin{table*}[t]
\begin{tabular}{|p{0.45cm}|p{1.25cm}|p{1.29cm}|p{1.5cm}|p{1.65cm}|p{1.4cm}|p{2.5cm}|p{1.6cm}|p{2.05cm}|}
  \hline
    Task  & Correction & Time (s) & $\delta$ GVF & $\delta$ Moran's I  & \# Classes & Class Methods & Colors & Projection  \\
  \hline
  1        & 98.3\%
     & $\mu=268, \newline \sigma=117$ & $\mu=0.11, \newline \sigma=0.03$ & $\mu=-0.01, \newline \sigma=0.04$ &
     $\mu=5.83, \newline \sigma=0.91$ &
     11 Fisher-Jenks \newline
     10 Jenks-Caspall \newline
     4 Maximum breaks \newline
     4 Quantile \newline
     1 Equal Intervals
     &
     12 greens \newline
     6 reds \newline
     6 viridis \newline
     3 YiGnBu \newline
     2 OrRd \newline
     1 RdBu*
     &
     9 Equal Earth \newline
     5 Albers \newline
     4 Natural Earth \newline
     1 Albers USA \newline
     \textcolor[HTML]{FF0000}{11 Mercator}
     \\
  \hline

  2   & 95.0\%
     & $\mu=223, \newline \sigma=130$ & $\mu=0.21, \newline \sigma=0.05$ & $\mu=-0.11, \newline \sigma=0.02$ &
     $\mu=5.57, \newline \sigma=1.07$ &
     27 Fisher-Jenks \newline
     3 Equal Intervals
     &
     15 viridis \newline
     12 greens \newline
     1 RdBu* \newline
     1 reds \newline
     1 YiGnBu
     &
     13 Equal Earth \newline
     5 Natural Earth \newline
     3 Albers USA \newline
     1 Albers \newline
     \textcolor[HTML]{FF0000}{8 Mercator}
     \\
  \hline
  3   & 100\%
     & $\mu=142, \newline \sigma=68$ & $\mu=0, \newline \sigma=0$ & $\mu=0, \newline \sigma=0$ &
     $\mu=6, \newline \sigma=0$ &
     30 Jenks-Caspall
     &
     30 reds
     &
     28 Albers USA \newline
     2 Albers
     \\
  \hline

\end{tabular}
\vspace{2mm}
\caption{An overview of the quantitative analysis of the experiment results. An asterisk next to the color scheme indicates a diverging palette. The number of participants choosing the Mercator projection is marked in red, due to strong recommendations against this kind of projection.}
\label{tab:summary_of_cases}
\vspace{-4mm}
\end{table*}

We conducted a validation study to gauge whether potential map designers can improve on flawed map designs. More specifically, the purpose of this study is to assess whether GeoLinter is effective at steering audiences away from cartographic blunders. The feedback from this study will inform the next iteration of GeoLinter (Section~\ref{sec:geolinter_v2}).
In our study, participants used the GeoLinter framework to fix flawed choropleth maps. Quantitative and qualitative measures were captured with respect to number of linting errors corrected and overall satisfaction with the framework.

\subsection{Participants}

We recruited 30 participants for the study (24 male, 6 female, $M_{age} = 27.5, SD_{age} = 4.21$). Recruitment targeted computer science students who had familiarity with programming, but no formal training in cartographic design. Participants were paid \$10 for the study, which took approximately 30 minutes including training for an effective rate of \$20/hr. 8 of the participants waived the payment.
The rationale for recruiting this demographic was to collect initial feedback from audiences with no formal training in cartography. We believe that this demographic is justified for a preliminary evaluation because we are not conducting a full-scale usability study. Instead, we are simply collecting feedback for the sole purpose of improving GeoLinter's next iteration.

\begin{figure}[!b]
\vspace{-3.9mm}
	\centering
	\includegraphics[width=1.0\columnwidth]{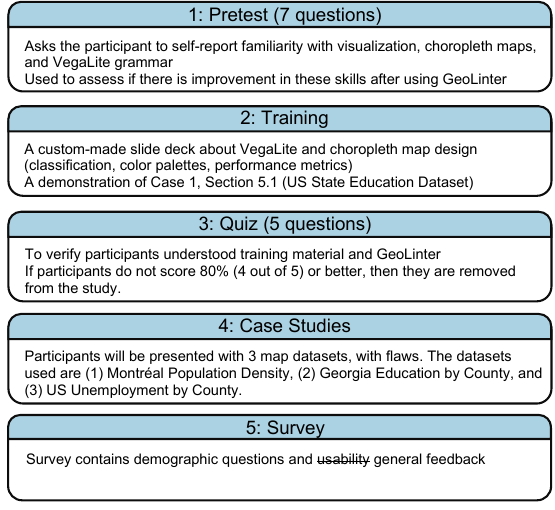}
	\vspace{-3.8mm}
	\caption{A summary of our validation study.} 
  \label{fig:experiment_design}
\end{figure}

\subsection{Procedure/Protocol}
Our study consists of five phases (as outlined in Figure~\ref{fig:experiment_design}): Pretest; Training; Quiz; Case Studies, and; Post-test and Survey.
The study begins with a pre-test questionnaire, gauging the participant's skill level and familiarity with choropleth maps and visualization grammars using a 5 point Likert scale (1 = Strongly Disagree, 2 = Slightly Disagree, 3 = Neutral, 4 = Slightly Agree, 5 = Strongly Agree). Participants averaged a score of 2.75 when asked ``I am familiar with reading choropleth maps.'' and 1.97 when asked ``I am familiar with creating choropleth maps.'', indicating that our pool of participants were not particularly confident in designing choropleth maps.

Next, a training session on VegaLite code format and choropleth map design principles and cartography is provided. Finally, each participant was given training on the GeoLinter Framework, walking through the dataset discussed in Case 1 (Section~\ref{subsection:case1_new}). At the end of the training session, the participants were given a quiz to verify that they have understood the material. Those that did not score above 80\% on the quiz were not considered in the study results. However, all participants who started the study passed. The participants were then asked to use GeoLinter to correct three designs (we refer to these as Tasks 1-3) with the objective: ``Design the map that most accurately reflects the underlying data.''

In each task, the participants were provided with an initial choropleth design with known flaws.
Task 1 contained the flaws \textcolor[HTML]{FFC000}{\textbf{LOW\_GVF}} and \textcolor[HTML]{FFC000}{\textbf{LEGEND\_COLOR}}.
Task 2 contained the flaws \textcolor[HTML]{FFC000}{\textbf{LOW\_GVF}} and \textcolor[HTML]{FFC000}{\textbf{BORDER\_COLOR}}.
Task 3 contained the flaw \textcolor[HTML]{FFC000}{\textbf{BG\_COLOR}}.
For all tasks, participants were given a full description of each dataset.

For each of the three tasks, we collected the soft rule correction rate, which was calculated by summing up all of the soft rule violations after the participant was finished with the task. We also collected the task completion time and the change between the presented design and final design with respect to GVF, Moran's I, number of classes, classification method, color palette, and projection method.
Results from the tasks are summarized in Table~\ref{tab:summary_of_cases}.
Finally, we conducted an exit survey, consisting of open-ended text responses as well as several repeated questions from the pre-test questionnaire to assess if participants gauged their skills differently after using GeoLinter.

\subsection{Results}
Results from our design corrections tasks showed a high adoption rate of the GeoLinter suggestions (greater than 95\%).
In Task 1, 29 out of the 30 participants corrected \textcolor[HTML]{FFC000}{\textbf{LEGEND\_COLOR}} and all of them corrected \textcolor[HTML]{FFC000}{\textbf{LOW\_GVF}}. Participants generally favored more classes, with $\mu = 5.83, \sigma=0.91$ for this task. GVF increased by 0.11 and Moran's I decreased by 0.01 on average. When selecting the color palette, 12 participants chose greens, 6 reds, 6 viridis, 6 YIGnBu, 2 OrRd, and 1 RdBu (diverging), indicating that the majority chose a green-related color scheme (the original design our participants were presented with used an unclassed viridis, the default color for VegaLite). The default map projection was Equal Earth, but the largest number of participants changed it to Mercator. None of the participants offered an explanation to this design choice. Most cartographers would agree that changing the projection from Equal Earth to Mercator is an objectively worse choice. That said, the usability study enabled us to identify a key weakness of the current system, the tested instantiation did not have the ability to guide participants toward the best projection for a map depicting a small territory with this shape at this latitude. Of the choices in the system, the best option is the Equal Earth, but that is only because a good choice is not included.

\begin{figure}[!t]
	\centering
	\includegraphics[width=1.0\columnwidth]{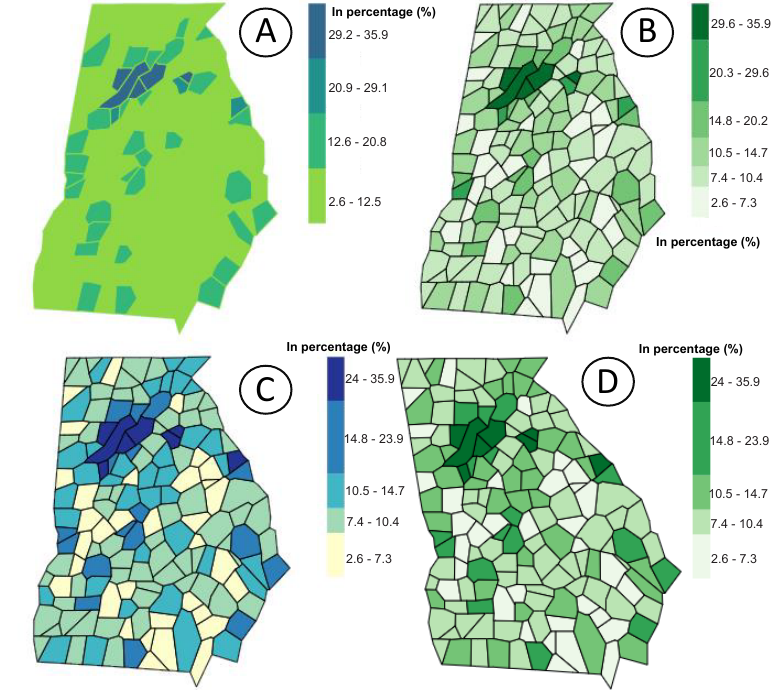}
	\caption{Exploring various design options to represent the county-level population distribution of the percent of residents with a bachelor's degree in Georgia, USA, 2016 (A) the choropleth map as defined by the original specification. (B) shows the design recommended by GeoLinter, which employs the equal earth projection. 
	(C) shows the design provided by participant 4 during our usability study, which uses a 5-class Fisher\addition{-}Jenks scheme with a YiGnBu palette under the Natural Earth projection. (D) shows the design provided by participant 17, which uses a 5-class Fisher\addition{-}Jenks under the Mercator projection.}
	\label{fig:case3_solutions}
	\vspace{-3mm}
\end{figure}


Task 2 uses the Georgia dataset shown in Figure~\ref{fig:case3_solutions}. 27 out of the 30 participants corrected \textcolor[HTML]{FFC000}{\textbf{BORDER\_COLOR}} and all of them corrected \textcolor[HTML]{FFC000}{\textbf{LOW\_GVF}}. Similar to Task 1, participants generally favored more classes, averaging 5.57 classes. GVF increased by 0.21 and Moran's I decreased by 0.11 on average. When selecting the color palette, 15 participants chose viridis, 12 greens, 1 RdBu (diverging), 1 reds, 1 YiGnBu, indicating that the majority kept the original scheme (the original design our participants were presented with was a viridis scheme). Once again, the default map projection was set at Equal Earth, with 8 participants changing it to the Mercator projection. Participant 17, who chose to use the Mercator projection said ``because that way it looks the closest to the layout of Georgia on a map.'' We further asked the participant what is Georgia supposed to look like, and the participant mentioned Google Maps, which uses the Mercator projection. Participant 17 lived in Alabama, directly west of Georgia, for 10  years. Participant 18 also offered an explanation as to why the Mercator projection was the best choice, stating that ``it distorts the poles ... for a state level map like this, it does not matter.'' Although partially valid, the explanation fails to capture the relative size distortion caused by the Mercator projection, which makes it generally less preferred by cartographers for choropleth mapping.
Figure~\ref{fig:case3_solutions} shows the design originally shown to the participants (A), the design if all of the GeoLinter corrections were adopted (B), and designs chosen from two participants (C-D).

Task 3 is a map of the United States showing the unemployment rate by county. In the initial design presented to participants, the legend was already optimized with a 6-class Jenks\addition{-}Caspall scheme with a GVF=$0.84$ and Moran's I$=0.68$. The only violation is that the background color is \#FC9272
\includegraphics[height=\fontcharht\font`\B]{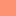}, which is the same as one of the fill colors in the legend (\textcolor[HTML]{FFC000}{\textbf{BG\_COLOR}}). All 30 participants corrected this issue. 28 participants selected the Albers USA projection, and 2 used Albers equal area conic projection.

Task 1 and 2 demonstrate that users tend to blindly follow GeoLinter suggestions, with some injecting their own flawed cartographic knowledge. While this behavior is understandable given the users' unfamiliarity with choropleth map design, the maps they created using GeoLinter's recommendations revealed flaws in the tool's first version. Specifically, the tool's use of compromise projections like Mercator is problematic, as cartographers argue that only equal-area projections are necessary. Additionally, many participants used the equal earth projection, which is not suitable for regional maps like U.S. states. As a result, there is a need to implement a feature for selecting appropriate equal-area projections.

Another issue identified was the lack of titles and units in the legends, which a cartographer expert confirmed as essential elements in all maps. This realization prompted consideration of an additional improvement - implementing checks to ensure that the data is appropriate for choropleth mapping. These suggested improvements are discussed in detail in Section~\ref{sec:geolinter_v2}.

We also asked each participant \textit{What is your overall impression of this tool?} with a free-form response field. We received positive feedback. 17 out of the 30 participants provided the brief response ``good''. 8 participants also said that GeoLinter was easy to use, for example ``It is very easy to use, even for someone with no previous experience.'' Another response was ``I feel this is a very useful tool to interactively develop choropleth maps. The linting suggestions that continually update as the map is interactively modified makes it very easy for a not very experienced designer to design maps that visually and aesthetically appealing in a nice guided manner.'' This comment captures the usability of GeoLinter.

We also asked the question \textit{What would you say about this tool’s effectiveness as a learning tool?} with a free-form response field. The response were also positive, for example, one participant said ``I could see it can prompt me where I could improve and what was problem with current visualization.'' One participant focused on the iterative nature of our pipeline, which was effective at repeatedly correcting errors until there are no more problems: ``The interactive nature and continuous updating makes it easy for novices to quickly gain feedback on their design choices when creating such maps. Additionally, the suggestions provided are clear and actionable: the users can understand what is wrong and what corrective steps to take to improve the map.''

In addition to the positive feedback, participants also provided suggestions for improvements to GeoLinter. We summarize these into three categories: metrics, color, and tutorials.

\vspace{1.5mm}
\noindent \textbf{Metrics.} In the middle of the study, participant 12 said that deciding the tradeoffs between GVF and Moran's I was difficult and explained that Moran's I was hard to grasp. Other participants also voiced their concerns about the lack of in-depth information about our performance measures: ``More information about Moran's I and GVF!'' and ``Current explanations are good enough, but maybe more that are specific to metrics.'' Adding more training or informative tooltips about performance measures can help expand the reach and accessibility of our framework, especially for novices.

\vspace{1.5mm}
\noindent \textbf{Color.} Three participants directly stated that there should be more color options. One participant also suggested limiting the options for the \addition{outline width}: ``Make stroke width and color options limited instead of allowing the user set their own.'' Incorporating research on color theory and drawing upon more of the research related to the perceptual effects of color (e.g. the further two map symbols are distanced, the more difficult it is to perform comparison tasks~\cite{brychtova2017effect}, larger color distances are beneficial for general choropleth map reading~\cite{brychtova2016empirical}, darker glyphs tend to be associated with higher values~\cite{mcgranaghan1989ordering}) on map data could greatly benefit future iterations of GeoLinter. The next extension of GeoLinter will include more help about the implications of different color choices and about which color schemes are essentially equally effective. This will enable the designer to pick the color scheme that best fits into their design goals (e.g., if the map is part of an organization report and that organization has particular hues in their logo and web site).

\vspace{1.5mm}
\noindent \textbf{Tutorials.} Six participants mentioned that more information, in general, should be added for each interface element. In the implementation of our interface, the narrative visual prompts and settings disappear after each soft rule is fixed. However, one participant suggested that we should keep the settings even after a rule is fixed: ``Improvements to make the UI more intuitive would be useful. Since I mostly just needed to click 'Fix', it was confusing when that button disappeared even though I still wanted to make changes.'' Another participant preferred if the VegaLite specification panel could be toggled: ``The JSON code used to generate the map could be hidden and optionally displayed if user desires to view and edit it.'' 



\section{GeoLinter v2}
\label{sec:geolinter_v2}

Our evaluation study findings (Section~\ref{sec:evaluation}) and reviews from the initial draft of this paper have been instrumental in identifying areas of improvement for GeoLinter. This led us to transition from our first iteration of the tool to an enhanced second version, which we discuss in detail in this section.

We observed that study participants, despite showing an inclination to adhere to GeoLinter's suggestions, sometimes opted for the Mercator projection contrary to the tool's recommendation for an equal-area projection. This trend indicates a possible gap between the tool's recommendations and user preferences or comprehension, hinting at the need for more comprehensive guidance or education within the tool about optimal projection choices for distinct map types and geographical extents. Furthermore, we identified a need for improvements in areas such as the inclusion of titles and units in the legends. These elements, essential in all maps as confirmed by our cartography expert, were not present in GeoLinter's initial iteration. We also realized that the importance of appropriate data selection for choropleth mapping had been overlooked. Building upon these insights, we present GeoLinter v2, showcasing enhancements derived from the evaluation study outcomes and a thorough literature review.

\subsection{Updated Map Considerations}
\label{subsection:updated_map_considerations}

In Section~\ref{sec:background}, we presented a review of relevant literature and the identified key considerations in map-making that were directly applicable to GeoLinter. These included guidelines derived from the literature for three key design decisions, those associated with (1) classification \addition{method}, (2) symbology, and (3) projection. In response to the significant frequency of user errors in the map projection selection, we revisited these guidelines and our implementation strategy for them to ensure more accurate results. Input from reviewers of the initial draft of this paper, a discussion with a cartographer expert on our research team, and a further analysis of Slocum et al.'s~\cite{slocum2014thematic} textbook on thematic map design suggested the inclusion of two additional critical design considerations: (4) title and legend design, and (5) data selection. This paper now presents the enhanced design constraints and goals integrating these important factors.

\vspace{1.3mm} \noindent \textbf{(C3) Map Projection:} Initially, we stated, "For small-scale maps much smaller than the scale of the Earth, conformal projections such as Mercator can be used, as long as the distortion of the area is minimal." However, following the validation study, we now only exclusively permit equal area projections. Moreover, we utilize Savric et al.'s~\cite{vsavrivc2016projection} Projection Wizard, which provides more suitable projection recommendations based on the type (e.g., equal-area, conformal, equidistant) and the extent (i.e., the size and shape of the area). For instance, an equal-area projection for regional maps in a square format, such as Wyoming, would employ an Oblique Lambert azimuthal equal-area. For a regional map with an east-west extent, such as Tennessee, an Albers equal-area conic would be utilized. In the case of a region with a north-south extent, like Vermont, a Transverse cylindrical equal-area projection would be adopted.

\vspace{1.3mm} \noindent \textbf{(C4) Legend Design:} 
As detailed in Slocum et al.~\cite{slocum2014thematic}, legend design can have a significant impact on the effectiveness of a map. The use of either a vertical or horizontal legend will depend on available map space. Numeric values should be placed in a consistent location, either at the bottom or to the right of legend boxes, and boxes should be arranged contiguously. Three methods for specifying class limits are: showing the actual data range in each class, expanding classes to eliminate gaps, or indicating minimum and maximum values and the upper limit for each class. A hyphen can be used to separate numeric values, and including a graphical display with the legend may be useful. The legend and units of the metric being displayed should also be clearly indicated, and should tie in with the title of the map. In GeoLinter, the default legend orientation is vertical, with all of the numerical labels automatically generated, such as in Fig.~\ref{fig:teaser}.

\vspace{1.3mm} \noindent \textbf{(C5) Selecting appropriate data}: 
According to Slocum et al.~\cite{slocum2014thematic}, the choropleth technique is most appropriate for phenomena that are uniformly distributed within each enumeration unit. However, in practice, this is seldom the case, and caution should be exercised in using choropleth maps. One issue to consider when selecting data for choropleth maps is to ensure that raw-total data have been adjusted to account for varying sizes of enumeration units. Four standardization approaches are discussed: 1) dividing an area-based raw total by some other area-based raw total, 2) creating a density measure by dividing a raw total not involving area by either the areas of enumeration units, 3) computing the ratio of two raw totals not involving area, and 4) computing a summary numerical measure (e.g., mean or standard deviation) for each enumeration unit.






Following the revisions to our design considerations, it was necessary to update the corresponding design goals accordingly. The alterations primarily encompass the inclusion and modification of design considerations C3, C4, and C5.

\vspace{1.5mm} \noindent \textbf{G2: Detect and explain suboptimal design parameters (C1, C2, C3, C4, C5).}

\vspace{1.5mm} \noindent \textbf{G3: Automatically fix the detected errors (C1, C2, C3, C4, C5).}

\subsection{Updated Framework}

\textbf{Map Projection}: 
The integration of Projection Wizard by Savric et al.~\cite{vsavrivc2016projection} helps in two ways: (1) assisting non-expert users in selecting an appropriate projection for their map, ensuring a more accurate representation of the spatial data; (2) offering a fully automatic solution that simplifies the process of choosing a suitable projection, thereby reducing potential errors and misunderstandings. The end result of this is fewer options for the designer, and increased reliability.

\vspace{1.5mm} \noindent \textbf{Selecting Appropriate Data:}
To facilitate the process of choosing suitable data for choropleth maps, we integrate data standardization tools into GeoLinter. These tools provide designers with options to standardize their data using the four approaches suggested by Slocum et al.~\cite{slocum2014thematic}. Through an interactive interface (Figure~\ref{fig:teaser}B and E2), designers can input their raw data, select the desired standardization method, and obtain the standardized data with proper units for use in their choropleth map. Maps generated from GeoLinter can only use accepted normalization techniques, further adding to its reliability.

\vspace{1.5mm} \noindent \textbf{Legend Design: } 
To assist designers in creating effective legends for their choropleth maps, GeoLinter incorporates the following features:

a. Legend orientation: GeoLinter allows designers to choose between vertical and horizontal orientations for their legends, providing a live preview of the changes as they make their selections.

b. Numeric value placement: GeoLinter enables designers to easily define the placement of numeric values in the legend, with options to display them either at the bottom or to the right of the legend boxes.

c. Legend and unit clarity: GeoLinter encourages designers to clearly indicate the legend and units in their map, tying them in with the map's title to ensure a cohesive design. To ensure that the title is descriptive, we employ a template to generate a generic title: [variable] in [region] over [time]. We also allow open-ended titles, such as the ones found in \textit{The Economist}, magazine where authors use a witty joke in lieu of a descriptive title.

 \section{Case Studies}
\label{sec:case}

\begin{figure}[!t]
	\centering
	\includegraphics[width=1\columnwidth]{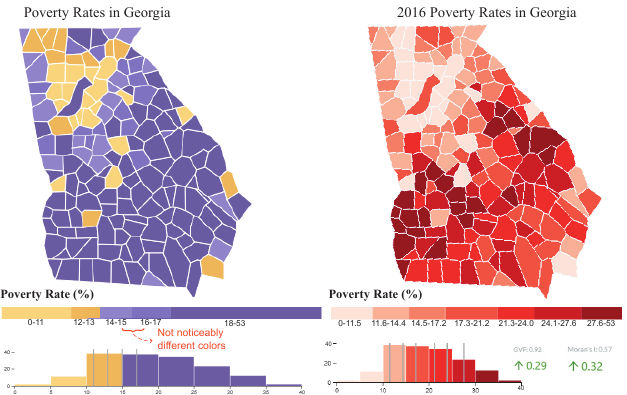}
	\vspace{-1.5mm}
	\caption{A published choropleth map showing the poverty rates in percentage in Georgia by county. (Left) The original design where the chosen classification method has a low GVF score and colors that are not easily separable. The bar immediately below it is the legend. The histogram at the bottom suggests that all values above 17 are in one class, and that more than half of all data values fall in this range. (Right) The proposed design based on the GeoLinter suggestions, along with the updated legend and histogram.}
	\label{fig:case2_new}
\end{figure}

In this section, we present case studies to demonstrate how our updated framework, GeoLinter v2, supports designers in creating choropleth maps.

\subsection{Case 1: Value of Freight Shipments by State 2002}
\label{subsection:case1_new}


In Figure~\ref{fig:teaser}, the designer is creating a choropleth map of US freight shipments in 2002. The Bureau of Transportation Statistics~\footnote{bts.gov/archive/publications/freight\_in\_america/figure\_13} published this data online, and has published its own version of a choropleth map (as shown in the Original Choropleth Map View in Figure~\ref{fig:teaser}B. The first violation is that the data is inappropriate for choropleth mapping (\textcolor[HTML]{FFC000}{\textbf{DATA\_NORM}}). The map shows absolute, un-normalized values. The designer chooses to normalize the data by dividing the value of freight shipments by the population to yield a derived, per-capita value.

\begin{figure*}[ht]
	\centering
	\includegraphics[width=1.0\linewidth]{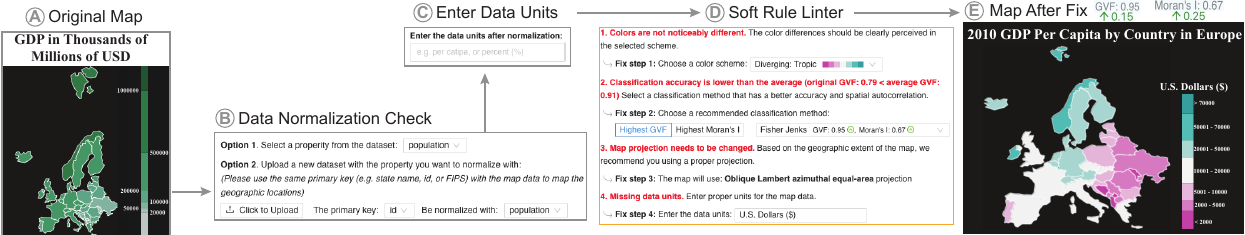}
	\vspace{-2.9mm}
  \caption{Case 3, featuring a dark UI map. After setting the original map (A) as the input, GeoLinter provides two options to normalize the data (B) and select the proper data units (C); then the Soft Rule Linter (D) notifies the designer of 3 potential flaws in the original map, specifically 1) not noticeably different colors, 2) low data classification accuracy, and 3) inappropriate map projection. GeoLinter produces (E) to address all errors.}
  \vspace{-4mm}
 \label{fig:case3_new}
\end{figure*}

The designer continues to fix design violations as detected by GeoLinter (Figure~\ref{fig:teaser}E). The next design error is the choice of colors in the legend. In the original map, the color palette contains two hues that are difficult to distinguish, thus triggering the \textcolor[HTML]{FFC000}{\textbf{LEGEND\_COLOR}} violation. Also,
the GVF score of the original design triggers the \textcolor[HTML]{FFC000}{\textbf{LOW\_GVF}} violation. For each of these issues, the Detected Violations panel (see Figure~\ref{fig:teaser}E) itemizes each soft rule violation along with an explanation on how to address each issue. The first step is to fix the legend fill colors that are too similar to each other, the designer selects a color palette defined by ColorBrewer~\cite{harrower2003colorbrewer}.
The histogram after normalization for this dataset shows that the values range from 0 to 50,000, with no significant skew. In order to address this issue, the designer may choose more unique classes and a color scheme with higher color variance to best show the full extent of the data. In this case, GeoLinter recommends a 7 class Jenks\addition{-}Caspall classification of the data. This scheme is used in order to capture the nuances of the nonuniform data distribution.

The interface continues to provide guidance on the map projection (Figure~\ref{fig:teaser}F.1), which is important in reducing the overall distortion of the map. GeoLinter recommends the Albers Equal Area projection, an equal-area projection that is appropriate given the geographic extent. No major problems with background and border stroke are identified. After following the on-screen prompts (Figure~\ref{fig:teaser} E.2, E.3, E.4), the designer creates a choropleth map using 7 classes with Jenks\addition{-}Caspall with a YiGnBu color palette on an Albers USA projection. The final design appears in the After Soft Fix View (Figure~\ref{fig:teaser}C), and the corresponding code (Figure~\ref{fig:teaser}A) is also automatically updated for embedding in websites.

\subsection{Case 2: Poverty Rates in Georgia, USA}
\label{subsection:case2_new}

The next case features a map created by Novoco~\footnote{novoco.com/atom/143541}, a national professional services and consulting organization. It published a map of poverty rates in Georgia (see Figure~\ref{fig:case2_new} left). At first, glance, this map is problematic, as most of Georgia is colored \#6E639B \includegraphics[height=\fontcharht\font`\B]{figures/colors/6e639b}, corresponding to the class that captures the data ranging from 17.1\%-52.2\%. This could potentially destroy meaningful information and insights that could otherwise be seen. When inspecting the histogram view (Figure~\ref{fig:case2_new}), the 17.1\%-52.2\% class appears to contain the majority of the data instances, which means the map can benefit from splitting the class into multiple classes.

In this case, GeoLinter recommends a 7-class quantiles classification \addition{method} with the sequential reds palette (Figure~\ref{fig:case2_new} right). We see that the map is able to portray the data with higher granularity; the areas that were previously colored \#6E639B \includegraphics[height=\fontcharht\font`\B]{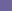} for  17.1\%-52.2\% are now split up into 4 classes. GeoLinter recommends an Oblique Lambert azimuthal equal-area projection. In this case study, GeoLinter may not explicitly flag soft rule violations, but can still guide the developer to an improved map design.



\subsection{Case 3: Dark UI: GDP by Country in Europe}
\label{subsection:case3_new}

Due to the popularity of dark UIs (such as in Google Chrome, Windows, etc.), we show how GeoLinter can also produce maps that work with dark UI backgrounds. We use a published map (Figure~\ref{fig:case3_new} (A)) of the GDP of Europe~\footnote{github.com/Nicknyr/Europe\_Choropleth}.

There are five rule violations in this case. The first rule violation is \textcolor[HTML]{FFC000}{\textbf{DATA\_NORM}}, as the map is showing the raw GDP of Europe. The designer chooses to address the rule violation by choosing to divide the GDP by the population of each country, producing a per capita value. The second rule violation, which is correcting an invalid title and legend with proper units (\textcolor[HTML]{FFC000}{\textbf{TITLE\_LEGEND}}) can be solved immediately after normalizing the data. The designer chooses to title the map ``2010 GDP per Capita by Country in Europe'', denoted by the U.S. dollar in the legend.

The third rule violation continues to improve the legend by selecting a valid color palette (\textcolor[HTML]{FFC000}{\textbf{LEGEND\_COLOR}}). Since this case study features a dark background, the designer chooses the Tropic diverging color scheme from CARTO~\footnote{carto.com/carto-colors} to separate above-average and below-average values. The fourth issue to fix is to the classification performance (\textcolor[HTML]{FFC000}{\textbf{LOW\_GVF}}). The designer chooses a 7-class Fisher\addition{-}Jenks to raise the GVF from 0.79 to 0.91, improving the classification accuracy. Finally, the last issue to address is the projection. The most appropriate equal-area projection for a map of Europe is the ``Oblique Lambert azimuthal equal-area,'' which is automatically selected. Based on the juxtaposition of the map, it is clear that the original map (Figure~\ref{fig:case3_new} (A)) had significant areal distortion. GeoLinter updates the VegaLite specification and presents the fixed map as Figure~\ref{fig:case3_new} (E) shows.





\section{Conclusion}
\label{sec:conclusion}
In this paper, we present GeoLinter, a framework for automatically detecting flaws in choropleth maps and recommending fixes for both hard and soft rules. The linter compiles a set of design guidelines and metrics from prior research in cartography to aid in selecting design parameters to enable the creation of choropleth maps that most accurately reflect the underlying data conform to well-established cartographic guidelines. Through a validation study, we identified prevalent cartographic errors often made by inexperienced map designers. These findings prompted us to enhance and refine GeoLinter, aiming to minimize the vulnerability to these specific errors. Although GeoLinter currently supports VegaLite, our framework can easily be extended to other visualization grammars. We believe it will have great use as an add-on in productivity and data visualization software. While most software tools rely on preprogrammed logic to determine recommended views for geographical data, sometimes the default settings can result in suboptimal layouts. Embedding our framework in Microsoft Excel and Tableau can help with creating both accurate and well-designed choropleth maps by presenting options and understandable explanations for critical design parameters.

We also believe that the GeoLinter framework could be beneficial for visualization designers who want to learn choropleth map making. A participant from our study directly stated ``This tool helped me learn a lot about how to make effective choropleth maps.'' Thus, our framework can serve as a educational tool that teaches the basic principles of choropleth mapmaking. Visualization designers can utilize our framework to test out different parameters in their designs. 

\vspace{1mm} \noindent \textbf{Limitations and Future Work: } We do note that GeoLinter is not without limitations. While GeoLinter embodies a robust system of checks to assist our target audience in avoiding cartographic blunders, one question that goes unanswered is whether a choropleth map the best type of visualization for the task at hand.  GeoLinter facilitates users in creating technically accurate maps adhering to established guidelines, however, the tool does not offer assistance in determining the optimal map choice. Thus, a future work or extension could include a function to determine the most fitting map type given the task and dataset.

Another limitation is the framework complexity. Several participants suggested adding tutorials, training, and more information on interface features and performance metrics. Even though some of the terminology and concepts introduced in GeoLinter require extended training, they cannot be eliminated due to its importance in choropleth map design. Future work will focus on developing simpler and more intuitive explanations on the more technical aspects of cartography for future iterations of a map linter, and we will adapt other strategies for building help into visualization tools that target non-experts (e.g.~\cite{kang2003helping}).

Our work and previous visualization linters~\cite{chen2021vizlinter,hopkins2020visualint} involve compiling guidelines developed by empirical research into a system. However, this approach may not cater to all audiences at all times. McNutt and Kindlmann~\cite{mcnutt2018linting} argued that linting should consider search or query intent. In Chen et al.'s study on VizLinter~\cite{chen2021vizlinter}, several participants rejected some of the rule recommended by their system, influenced by their own preferences or previous experience in visualization, which we also observed in our framework, primarily relating to the choice of projection. A future direction for our work could involve manual specification of intended goals in order to better match audience expectations.
We also plan to extend the rule base and functionality of GeoLinter beyond basic choropleth maps. The current version of GeoLinter only contains 4 encoding errors and 6 soft rules, sufficient to address the issues of most static, univariate choropleth maps. As mentioned in Section~\ref{sec:framework}, each rule is evaluated independently. However, in practice, changing one design parameter affects others as well, and there are multiple solutions for a single violation. Accounting for these interactions should be done in future work in this direction.

Finally, a significant future direction for our work involves conducting comprehensive evaluations of the GeoLinter framework. This would entail submitting the maps produced by GeoLinter to a panel of experts in visualization and cartography for assessment. This approach would not only provide an in-depth understanding of GeoLinter's performance, but also highlight potential areas for further refinement and enhancement of the system.




\ifCLASSOPTIONcompsoc
  \section*{Acknowledgments}
\else
  \section*{Acknowledgment}
\fi

This work was supported in part by the National Science Foundation under Grant 1639227 and the U.S. Department of Homeland Security under Award 2017-ST-061-QA0001.

\ifCLASSOPTIONcaptionsoff
  \newpage
\fi



\bibliographystyle{IEEEtran}
%


\bibliography{template}

%




\begin{IEEEbiographynophoto}{Fan Lei}
is a Ph.D. student at Arizona State University in the School of Computing and Augmented Intelligence. He received his masters degree in Computer Science from Durham University, UK. His broad research interests include visual analytics, spatio-temporal analysis, information visualization, human-computer interaction, and explainable artificial intelligence.
\end{IEEEbiographynophoto}

\vspace*{-1cm}

\begin{IEEEbiographynophoto}{Arlen Fan}
is a Ph.D. student at Arizona in the School of Computing and Augmented Intelligence. He received his B.S. in Electrical and Computer Engineering from the University of Rochester in Rochester, New York. His research interests include visualization design focusing on addressing design errors and improving communication.
\end{IEEEbiographynophoto}

\vspace*{-1cm}

\begin{IEEEbiographynophoto}{Alan M. MacEachren}
is a professor of geography, affiliate professor of information sciences \& technology, and director of the GeoVISTA Center (www.GeoVISTA.psu.edu) at Pennsylvania State University. He is PI for the Penn State component of the DHS VACCINE Center of Excellence. His research foci include: geovisual analytics, geovisualization, geocollaboration, spatial cognition, human-centered systems, and user-centered design. He was an associate editor of IEEE Transactions on Visualization and Computer Graphics (TVCG) (2007-2011).
\end{IEEEbiographynophoto}

\vspace*{-1cm}

\begin{IEEEbiographynophoto}{Ross Maciejewski}
is a professor at Arizona State University in the School of Computing and Augmented Intelligence and Director of the Center for Accelerating Operational Efficiency - a Department of Homeland Security Center of Excellence. His primary research interests are in the areas of geographical visualization and visual analytics focusing on homeland security, public health, dietary analysis, social media, criminal incident reports, and the food-energy-water nexus. He has served on the organizing committees for the IEEE Conference on Visual Analytics Science and Technology and the IEEE/VGTC EuroVis Conference.
\end{IEEEbiographynophoto}


\vfill




\end{document}